# Magnetic anisotropic pinning and symmetric breaking induced by interfacial coupling in topological-like ruthenate superlattices

Zhongyuan Jiang[1,#], Zhiwei Zhang[2,#], Kesen Zhao[3,#], Wenjie Meng[3], Yuanyuan Zhao[2], Yubin Hou[3], Zhangzhang Cui[1], Jian Zhang[1], Zheling Shan[1], Haoliang Huang[2*], Qingyou Lu[1,3*] and Yalin Lu[1*]

[1]*Hefei National Laboratory, Anhui Laboratory of Advanced Photon Science and Technology, Hefei National Laboratory for Physical Sciences at the Microscale, University of Science and Technology of China, Hefei 230026, China*
[2]*Quantum Science Center of Guangdong-Hong Kong-Macao Greater Bay Area, Shenzhen 518045, China*
[3]*Anhui Province Key Laboratory of Condensed Matter Physics at Extreme Conditions, High Magnetic Field Laboratory, Chinese Academy of Sciences, Hefei, Anhui, 230031, China*

#Authors contributed equally.

**Correspondence to:** Dr. Haoliang Huang, Quantum Science Center of Guangdong-Hong Kong-Macao Greater Bay Area, Shenzhen 518045, China. E-mail: huanghaoliang@quantumsc.cn; ORCID: 0000-0002-5686-5519. Pro. Qingyou Lu, Anhui Province Key Laboratory of Condensed Matter Physics at Extreme Conditions, High Magnetic Field Laboratory, Chinese Academy of Sciences, Hefei, Anhui, 230031, China. E-mail: qxl@ustc.edu.cn; ORCID: 0000-0003-1934-8165. Prof. Yalin Lu, Hefei National Laboratory, Anhui Laboratory of Advanced Photon Science and Technology, Hefei National Laboratory for Physical Sciences at the Microscale, University of Science and Technology of China, Hefei 230026, China. E-mail: yllu@ustc.edu.cn; ORCID: 0000-0001-8240-5404.

## Abstract

Interfacial engineering enables various emergent effects such as spin reorientations and transport anisotropy. Noncollinear spin textures are essential for realizing many emergent quantum transport phenomena. However, driving such spin structures requires precise control of the interfacial magnetic coupling in complex oxide heterostructures. Here, by utilizing competing exchange interactions at the interface between ferromagnetic metal $SrRuO_3$ and ferromagnetic insulator $LaCoO_3$, we discovered a noncollinear spin configuration in $SrRuO_3$ sublayers. Magnetic stripes were induced by out-of-plane rather than in-plane magnetic fields, indicating strong anisotropy pinning in our superlattices. The observed magneto-transport anisotropy is well explained by our proposed spin configurations, accounting for contributions from both bulk and interface of the $SrRuO_3$ layers. More interestingly, magnetic skymionic textures were absent even at high magnetic fields. The interfacial exchange interaction overwhelms the Dzyaloshinskii-Moriya interaction (DMI) that stabilizes skyrmions, featuring a higher exchange coupling energy than that for the topological spin textures. Our work highlights the potential of interfacial engineering in tuning the spintronic properties by designing proper interfacial interactions.

# INTRODUCTION

Interfacial engineering in complex correlated oxide heterostructures has emerged as a powerful paradigm for designing novel quantum states of matter. At atomic-layer-thick interfaces, the appearances of lattice strain [1–4], electronic reconstruction [5], superconducting pairing [6–8] and spin reorientation [9] can give rise to various emergent phenomena not found in the constituent bulk materials. These engineered quantum phases hold significant potential for next-generation electronic and quantum information technologies. For example, customized interface exchange coupling can be achieved by delicately regulating the selection of the bulk material, growth conditions as well as design of the inter-layers, thereby obtaining spintronic functions designed on demand [1,4,10]. In particular, when two materials with distinct magnetic orders form an interface, the mismatch of spin configurations at the interface will induce significant magnetic proximity effects and lead to abnormal electronic transport characteristics [11–13], providing an ideal platform for exploring novel quantum phenomena.

SrRuO$_3$ (SRO), for instance, is a well-known itinerant ferromagnet ($T_C$~120 K) with unique transport and magnetic properties. SRO-based interfaces can host multiple interesting effects, such as interfacial oxygen distortion, strain-induced anisotropy, anomalous Hall effect (AHE), and skyrmionic spin textures, offering potential routes to design switches of various spintronic behaviors [2,14–23]. It is worth noting that the specifically designed SRO heterostructures exhibit hump-shape Hall resistivity arising at particular field strengths and temperatures, which is sometimes alternatively attributed to the topological Hall effect (THE) associated with magnetic skyrmions [24,25], leaving the origin still highly debated [26–28]. Furthermore, SRO can exhibit perpendicular magnetic anisotropy (PMA) under compressive strain with the spin orientation preferring the film normal direction [29–32]. In contrast, while a non-magnetic insulator ground state in bulk, tensile-strained LaCoO$_3$ (LCO) thin films can stabilize a ferromagnetic (FM) insulating phase with a lower Curie temperature (75~87 K) and exhibit in-plane magnetic anisotropy (IMA) [33,34]. This stark contrast, i.e. a strained FM metal with PMA (SRO) adjacent to a strained FM insulator with IMA (LCO), creates an ideal interface for generating profound magnetic frustration and symmetry breaking. The direct exchange coupling at such an interface is expected to compete with the intrinsic anisotropy of each layer, potentially leading to noncollinear spin reorientation, exotic domain structures, and manipulation of topological defects.

In this study, a series of high-quality LCO/SRO superlattices with varying individual layer thicknesses were prepared. We proposed a robust noncollinear ferromagnetic spin reoriented configuration induced at the interfaces between LCO and SRO layers with the distinct magnetic properties. This configuration suppresses the topological Hall signature typically associated with skyrmions in SRO. Instead, we observe the emergence of highly anisotropic, stripe-like magnetic domains that are indicative of a nematic magnetic phase. The detailed symmetry analysis of the magneto-transport as a function of temperature and magnetic field direction allows us to map the temperature- and field-driven evolution of the spin orientations. We attribute the formation of these magnetic stripes to a strong, anisotropic exchange interaction at the LCO/SRO interfaces, which largely screens the Dzyaloshinskii-Moriya interaction (DMI) responsible for skyrmion stabilization while imposing a directional preference on the magnetic order. This work demonstrates how interfacial design can be used to suppress one topological texture in favor of another potentially useful nematic state, highlighting a new pathway for quantum material engineering.

# MATERIALS AND METHODS

**Thin film deposition and superlattice preparation**

The superlattices (LCO$_M$/SRO$_N$)$_{12}$ (denoted by L$_M$S$_N$) were grown on TiO$_2$-terminated SrTiO$_3$ (STO) substrates (from MTI Corporation Inc., China) by pulsed laser deposition (PLD) assisted with reflection high energy electron diffraction (RHEED). We have successfully designed and prepared various superlattices, among which L$_9$S$_6$, L$_7$S$_6$, and L$_5$S$_5$ were particularly investigated. During the growth of the superlattices, the substrate temperature, oxygen partial pressure, and repetition rate of the laser pulse were 750°C, 100 mTorr, and 2 Hz, respectively. The laser fluence set at ~2 J cm$^{-2}$ for the SRO layer and 1.2 J cm$^{-2}$ for the LCO layer.

**X-ray Characterizations and Scanning Tunneling Electron Microscopy**

X-ray characterizations, including X-ray diffraction (XRD), X-ray reflectivity (XRR), and reciprocal space mapping (RSM), of the superlattices were performed using a four-circle diffractometer (Rigaku, SmartLab, Cu Kα). Scanning Tunneling Electron Microscopy (STEM) were performed using an FEI Titan Themis G2, equipped with a double spherical-aberration corrector and a high brightness field-emission gun. The cross-section of the specimens was prepared using a FEI Helios 600i dual-beam FIB/SEM machine. Pt electron-beam deposition and C ion-beam deposition were conducted in order to protect the sample surface from beam damage. Both X-ray characterization and STEM experiments were conducted at room temperature.

**Magnetic Force Microscopy**

Magnetic Force Microscopy (MFM) were performed in a 12 T superconducting magnet placed in ultrahigh vacuum at the High Magnetic Field Laboratory in Hefei, China. The spatial resolution of the microscope is 10nm. A rotation sample plate is mounted on the axis, allowing for an *in-situ* field angle ranging from 0° to 180°. For signal detection, we utilized custom piezoresistive cantilevers (SCL-Sensor Tech., PRSA-L300-F50-STD) characterized by a hollow beam structure and a resonance frequency of ~37 kHz. The tips were magnetically coated with a Ti (5 nm)/Co (30 nm)/Au (5 nm) film using electron beam evaporation and pre-magnetized perpendicular to the cantilever beam. The readout electronics consisted of a home-made active Wheatstone bridge preamplifier (gain 1000×) interfaced with a commercial RHK R9 controller. Topographic images were acquired in tapping mode, while MFM images were acquired in constant-height mode approximately 100 nm above the sample surface. The resulting MFM contrast maps the resonant frequency shift caused by the out-of-plane stray field gradient, where dark and light regions represent attractive and repulsive magnetic forces, respectively. Post-processing of the SPM data was conducted using WSxM.

**Magnetic-field-angle-dependent resistivity**

All transport measurements were performed using a physical property measurement system (PPMS, Quantum Design) equipped with a sample rotator. The magnetic-field-angle-dependent resistivity (MAR) is defined as MAR = $[\rho(\theta) - \rho(0°)]/\rho(0°)$, where $\theta$ is the angle of the magnetic field to film normal (0˚). The absolute MAR (AMAR) is defined as the absolute value of MAR and is suitable for exploring rotational symmetry from a butterfly plot of AMAR vs. $\theta$.

# RESULTS AND DISCUSSION

**Layer-by-layer quality at the atomic level**

LCO/SRO superlattices were prepared by alternating growth of LCO and SRO on top of TiO$_2$-terminated STO substrates. **Figure 1a** shows the conceptual stacking of the LCO and the SRO layers on STO. The STEM image identifies each atomic layer with no ambiguity. RHEED images during sample preparation were captured for the three sets of samples (**Figure 1b–d**), which present the layer-by-layer nature of the growth. XRD measurements (**Figure 1e**) show rigid epitaxy of the superlattice to the substrate according to the result of the (002) substrate peak and the superlattice satellite peaks (SL0, ±1, ±2), including beating patterns due to the whole film thickness and the satellite peaks (SLx) due to each LCO/SRO layer pair. The XRR patterns were in good agreement with the theoretical fit for L$_9$S$_6$, L$_7$S$_6$ and L$_5$S$_5$ structures, respectively (**Figure 1f**). The combination of RHEED, STEM, XRD and XRR shows the high atomic layer-by-layer quality for all of our superlattice samples.

**Magnetic anisotropic pinning**

We performed the cryogenic MFM measurements to directly image the magnetic domain evolution for our superlattice samples. Here, we focus on the L$_9$S$_6$ samples. **Figure 2a** shows a series of the MFM images at 75 K with the field applied perpendicular to the film plane and swept in a complete loop. Magnetic stripe domains emerge at ~1 T, become suppressed at higher fields (> -10 T), and reappear as the field decreases back to -1 T from negative saturation. Same evolution occurs as the field continues to increase to 10 T and returns to zero. This indicates that the magnetic stripe domains are stabilized at intermediate fields (~1 T), but disperse at both low fields (<0.5 T, near demagnetized state) and high fields (>10 T, field-polarized state). **Figure 2b** shows corresponding MFM images at 25 K. A notable difference is that at lower temperature, stripe domains appear at smaller field strength (~0.3 T) and persist to higher fields, indicating enhanced magnetic anisotropy pinning. Complete sets of MFM loop measurements with out-of-plane fields can be found in Supplementary Figures S1–S4. By contrast, when the field is applied in-plane, no magnetic stripe is observed even at 10 T (See Supplementary Figures S5–S6), demonstrating that the magnetic stripe formation is highly anisotropic and favored by out-of-plane magnetic fields. While magnetic anisotropy has been reported in similar materials [35–37], the stripe morphology here is exclusive to our superlattice architecture. The field-induced suppression of stripes at high out-of-plane fields (>8 T) reflects a transition toward a uniformly magnetized state.

**Symmetry breaking by field and temperature**

To further investigate the magnetic anisotropy, we conducted the MAR measurements as a function of temperature and magnetic field. Butterfly plots of the AMAR are presented in **Figure 3** for the $L_9S_6$ samples, where $\theta$ is defined as the angle between the applied field and the film normal. Given that the bulk LCO is insulating, the measured conduction would exclusively come from bulk SRO and SRO/LCO interfaces. A purely two-fold symmetric AMAR signal would be expected if the SRO magnetization maintains a single, fixed orientation. Therefore, the emergence of any higher-order symmetry component must originate from intrinsic spin reorientations within the SRO layers. Strikingly, our data reveal a distinct four-fold symmetric component coexisting with the fundamental two-fold symmetry (**Figures 3i–l**). Comparing data by rows, for example **Figures 3a–c**, the order parameter for the four-fold symmetry strengthens as temperature increases. Conversely, comparing data by columns, for example **Figures 3a**, **3d**, **3g** and **3j**, the four-fold component weakens as the magnetic field increases. Hysteretic behavior, indicative of coercive ferromagnetism, is observed but diminishes as higher temperature or field increases thermal fluctuations and promotes magnetization alignment.

Two key questions are posed by the AMAR results: i) the origin of the four-fold symmetry and ii) the unalignment of the minimum resistivity of the four-fold and two-fold components. To address these questions, we extracted the evolution of the field angle $\theta_{min}$ for the minimum resistivity across all measured conditions in **Figure 3** and summarized in **Figure 4a**. We find that $\theta_{min}$ increases with both field strength and temperature. Since bulk LCO and SRO exhibits IMA and PMA, with in-plane and out-of-plane spin orientation, respectively, the observed transport anisotropy necessitates that the itinerant SRO adopts spatially varying spin reorientations. We therefore propose a spatially evolving spin texture within the SRO layers, driven by interfacial exchange coupling with LCO (**Figure 4b**). The strong interfacial coupling forces SRO spins adjacent to the interface to cant toward the in-plane direction (inheriting the IMA of LCO), while SRO spins in the layer interior retain a stronger tendency toward the out-of-plane orientation (reflecting the PMA of bulk SRO). The application of a magnetic field progressively aligns the interior out-of-plane spin component toward the in-plane direction, thereby degrading the four-fold symmetric order and eventually restoring a pure two-fold symmetry at high fields, consistent with our observations. More interestingly, higher temperature requires higher field to suppress the four-fold component (**Figures 3j**, **3k** and **3l**), indicating that the spin reoriented textures of SRO are more robust at high temperatures. This trend suggests that thermal fluctuations partially relieve the interfacial PMA-IMA competition, strengthening the noncollinear spin configuration against field-induced alignment.

This depth-dependent noncollinear spin-reorientation is confirmed by the AMAR butterfly plots shown in **Figure 3**, but remains unresolved the correlation to the anisotropy of the magnetic stripes. **Figure 4c** shows the MAR versus $\theta$ plot for various fields at 25 K. The minimal resistivity occurs at $\theta=90°$ (in-plane field), where no magnetic stripe is formed. We propose that the stripes formation introduces localized exchange interaction and enhanced scattering, thereby suppression of conductivity. More details about the MAR versus $\theta$ plots can be found in Supplementary Figure S7. Similar MAR behaviors have been reported in other SRO superlattice systems, where eight-fold symmetry was claimed [38]. Furthermore, the persistence of four-fold symmetry even when LCO approaches its Curie temperature (**Figures 3c, 3f, 3i** and **3l**) indicates robust interfacial coupling between LCO and SRO.

**Spin orientation and topological transport properties**

One may be curious about whether the LCO/SRO superlattices exhibit topological Hall features associated with skyrmion spin textures. We state that our MFM results found no signature of any skyrmion structure, which would typically appear as nanoscale circular spots. The suppression of topological skyrmions indicates that the DMI is overwhelmed by a stronger interfacial exchange interaction. However, subtle deviations from conventional anomalous Hall behavior are evident in our Hall resistivity measurements. **Figure 5** shows field-dependent Hall resistivity (after subtracting the ordinary Hall contribution) at 25 K with vertical magnetic fields, overlaid with MFM images at selected fields. Green symbols represent experimental data, and the red curve shows a fit using the conventional two-channel AHE model [27,28,39]. The fitting quality is reasonable except near the peak and valley regions ($\sim\pm1.2$ T), where significant deviations occur (See Supplementary Figure S8).

Strikingly, the MFM images reveal an anticorrelation between stripe-domain visibility and the Hall resistivity anomaly: stripe domains are pronounced at low fields ($-0.05$ T, $-1.0$ T) and high fields ($> -5$ T), but become significantly suppressed near $-1.2$ T—precisely where the Hall resistivity exhibits its peak feature. This anticorrelation demonstrates that the Hall anomaly does not originate from conventional domain-wall scattering (which would maximize when domain walls are most numerous). Instead, the peak structure reflects spin-reorientation-induced transport anisotropy: near $-1.2$ T, the competing interfacial and bulk anisotropies drive a transition from the noncollinear, stripe-domain state toward a more uniformly canted spin configuration, producing enhanced spin-chirality-dependent scattering that is not captured by the two-channel AHE model.

The two-channel AHE model assumes isotropic magnetic properties and thus provides no information about the detailed spin configuration in our samples. The systematic underfitting of peak and valley regions indicates residual contributions from the noncollinear spin texture—distinct from topological Hall effects associated with skyrmions, but nevertheless reflecting the complex interplay between spin-orbit coupling and the depth-dependent spin reorientation in our superlattices. Our combined MFM and MAR measurements thus provide critical constraints on the spin configuration, revealing how interfacial engineering can stabilize novel magnetic textures that suppress topological skyrmions while generating distinctive magneto-transport signatures through noncollinear spin arrangements.

## CONCLUSIONS

We synthesized LCO/SRO superlattices on STO substrates with various combinations of layer thicknesses using PLD. Highly anisotropic, one-dimensional magnetic stripes are induced by out-of-plane magnetic field at intermediate strengths ($\sim 1$ T), but are absent for in-plane field even up to 10 T, indicating a strong magnetic anisotropy pinning in these superlattice samples. Field- and temperature-dependent magneto-transport measurements show that these magnetic stripes domains are associated with elimination of conductivity and exhibits complex rotational symmetries that evolve systematically with experimental conditions. We proposed a depth-dependent noncollinear spin configuration within the SRO layers, where spins reorient from out-of-plane in the bulk interior to in-plane near the interfaces due to strong exchange coupling with LCO. This spatially varying spin texture quantitatively explains the observed four-fold symmetric magneto-resistance and its field- and temperature-dependent evolution. Notably, the four-fold symmetry persists even near the LCO Curie temperature, evidencing robust interfacial coupling. Significantly, no skyrmion structures are observed in either field direction up to 10 T. The suppression of topological skyrmions indicates that interfacial exchange interactions overwhelm the DMI that would otherwise stabilize chiral spin textures. Instead, the competition between interfacial and bulk magnetic anisotropies stabilizes the stripe-domain state. Our results reveal that by delicate interfacial engineering, various emergent magnetic effects such as topological Hall effect and magnetic anisotropy can be precisely tuned to allow for potential spintronic devices for quantum information applications.

## DECLARATIONS

### Authors' contributions

Experimental design: Haoliang Huang, Yalin Lu
Experiments and data collection: Zhongyuan Jiang, Kesen Zhao, Wenjie Meng, Zhangzhang Cui, Jian Zhang, Zheling Shan, Qingyou Lu
Data analysis: Zhongyuan Jiang, Kesen Zhao, Wenjie Meng, Zhiwei Zhang, Yuanyuan Zhao
Manuscript writing and revision: Zhiwei Zhang, Haoliang Huang, Yalin Lu
Supervision: Yalin Lu

### Availability of data and materials

The data that support the findings of this study are available from the corresponding author upon reasonable request.

### Financial support and sponsorship

This work was supported by the Natural Science Foundation of Anhui Province (No. 2208085ME113), Hefei National Lab (ZB2025020100), Guangdong Provincial Quantum Science Strategic Initiative (GDZX2501001, GDZX2401003 and GDZX2401004), National Key R&D Program of China (Grant 2023YFA1607701 and 2024YFA1408101), National Natural Science Foundation of China (Grants 51627901 and 12527803) and the support from International Station of Quantum Materials. We thank the staff members of the SMA System (https://cstr.cn/31125.02.SHMFF.SM2.SMA) at the Steady High Magnetic Field Facility, CAS (https://cstr.cn/31125.02.SHMFF), for providing technical support and assistance in data collection and analysis.

### Conflicts of interest

All authors declared that there are no conflicts of interest.

### Supplementary Materials

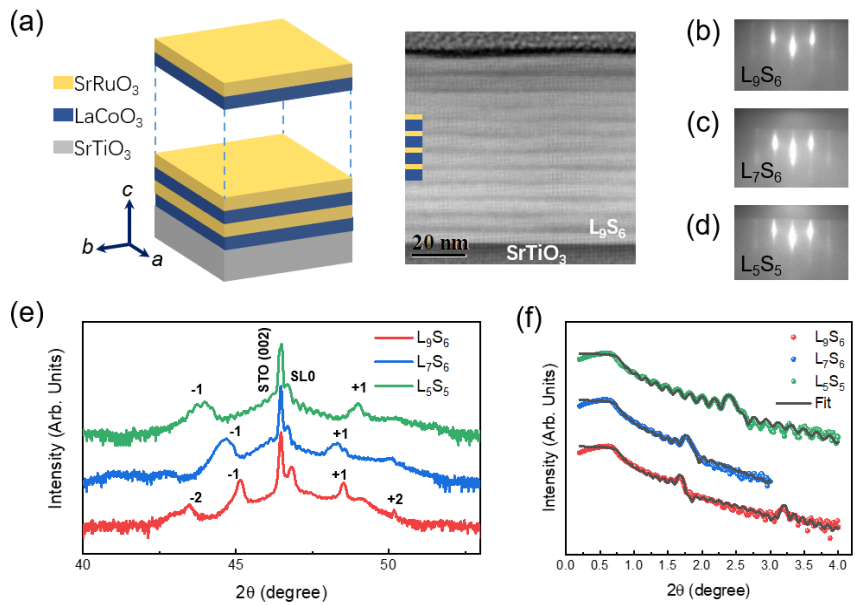

**Figure 1** Structural characterization of LCO/SRO superlattices. (a) Schematic illustration of the superlattice structure (left) and cross-sectional STEM image of the $L_9S_6$ sample (right), showing atomically sharp interfaces between LCO (blue) and SRO (yellow) layers grown on STO substrate (gray). The scale bar represents 20 nm. (b-d) RHEED patterns along the [100] azimuth for $L_9S_6$, $L_7S_6$, and $L_5S_5$, respectviely, exhibiting bright streaks indicative of layer-by-layer growth and atomically smooth surfaces. (e) XRD $\theta$-$2\theta$ scans around the (002) reflection, showing the substrate peak (STO), the superlattice zero-order peak (SL0), and satellite peaks ($\pm 1$, $\pm 2$) arising from the artificial periodicity. The distinct satellite peaks confirm the high structural quality and coherent epitaxial growth of the superlattices. (f) XRR data (symbols) and theoretical fits (solid lines) for the three superlattice samples. The well-defined Kiessig oscillations demonstrate uniform film thickness and sharp interfaces, with excellent agreement between experimental data and fits validating the designed layer thicknesses.

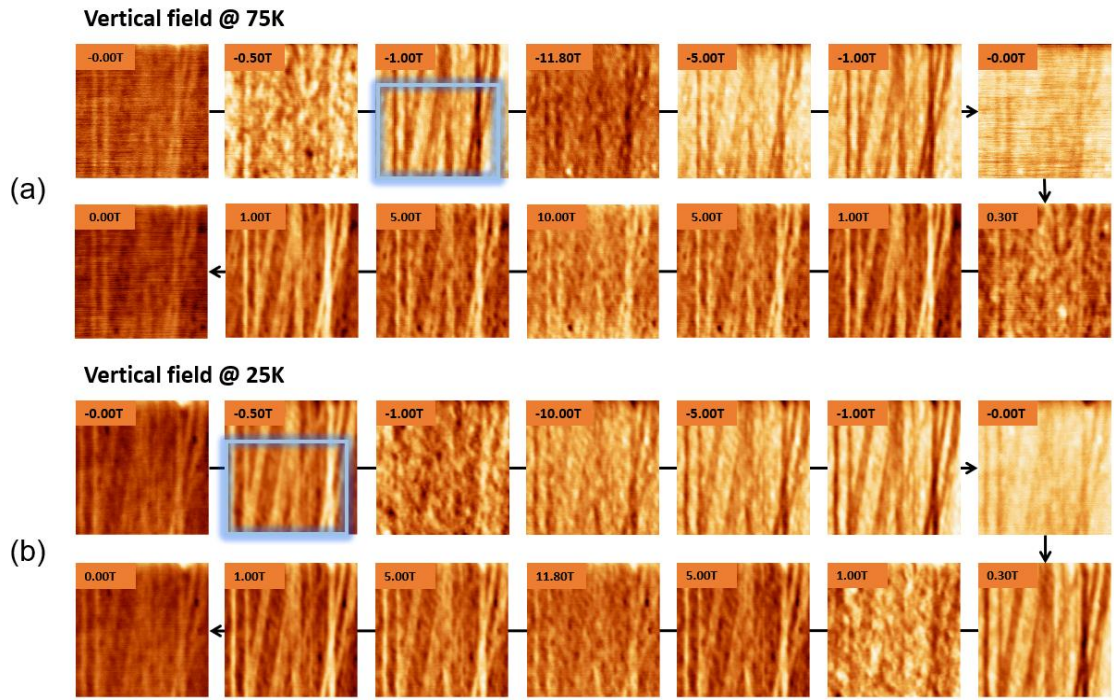

**Figure 2** Magnetic field-dependent MFM images of $L_9S_6$ superlattice under out-of-plane magnetic fields. (a) MFM images acquired at 75 K during a magnetic field loop applied along the out-of-plane direction. The magnetic field sequence is: −0.00 T → −0.50 T → −1.00 T → −11.80 T → −5.00 T → −1.00 T → −0.00 T → 0.30 T → 1.00 T → 5.00 T → 10.00 T → 5.00 T → 1.00 T → 0.00 T. The blue box highlights the field region (−1.00 T) where stripe domains exhibit optimal contrast. (b) Corresponding MFM images at 25 K with the field sequence: −0.00 T → −0.50 T → −1.00 T → −10.00 T → −5.00 T → −1.00 T → −0.00 T → 0.30 T → 1.00 T → 5.00 T → 11.80 T → 5.00 T → 1.00 T → 0.00 T. Arrows indicate the field sweep directions. At both temperatures, stripe-like magnetic domains emerge at intermediate fields (~1 T), become suppressed at high fields (>10 T), and reappear upon field reduction, demonstrating reversible field-driven magnetic transitions. Notably, the stripe contrast is more pronounced at lower temperatures and lower field strengths, indicating enhanced magnetic anisotropy pinning at 25 K. No circular skyrmion-like features are observed even at fields up to 11.8 T.

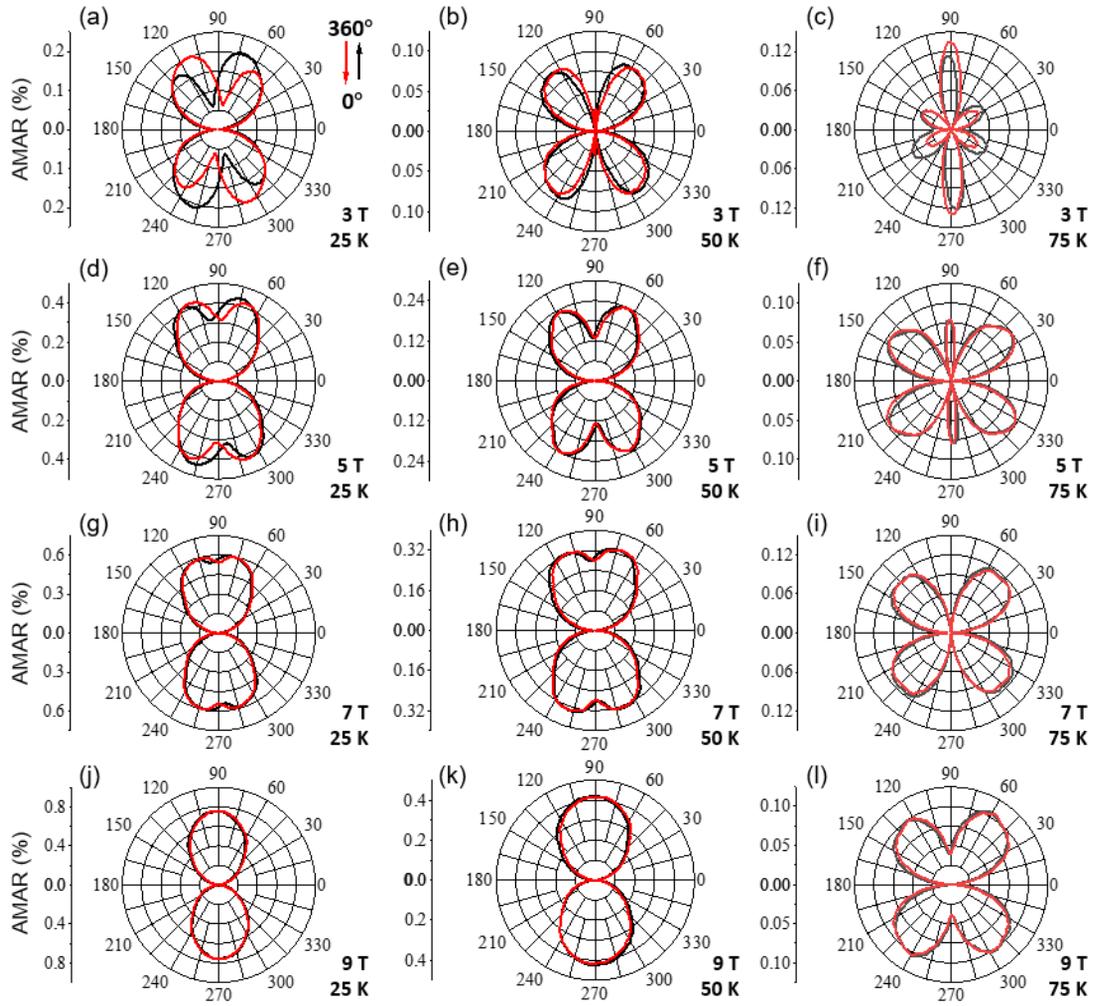

**Figure 3** Angular-dependent magnetoresistance (AMAR) butterfly plots of $L_9S_6$ superlattice. Polar plots of AMAR as a function of magnetic field angle $\theta$ at various temperatures and magnetic field strengths. $\theta = 0°$ corresponds to the magnetic field applied perpendicular to the film plane (film-normal direction), while $\theta = 90°$ and $270°$ correspond to in-plane field orientations. The radial axis represents the AMAR magnitude in percentage. Red curves show data acquired during counter-clockwise field rotation (360° → 0°), while black curves represent clockwise rotation (0° → 360°). The hysteretic behavior between red and black curves indicates magnetic anisotropy and irreversible spin reorientation processes. (a-c) AMAR at 3 T for 25 K, 50 K, and 75 K, respectively. (d-f) Corresponding data at 5 T. (g-i) Data at 7 T. (j-l) Data at 9 T. At 25 K and low fields (3-5 T), a pronounced four-fold symmetric pattern dominates, indicating a noncollinear spin configuration with contributions from both bulk and interfacial SRO layers. As the field increases to 9 T, the symmetry evolves toward a two-fold pattern, suggesting field-driven spin reorientation toward the film-normal direction. At 75 K, closer to the Curie temperature of LCO, the four-fold symmetry persists even at higher fields, evidencing robust interfacial exchange coupling. The coexistence of two-fold and four-fold symmetry components is most evident at intermediate temperatures (50 K) and fields (5−7 T), as shown in panels (e) and (h). The hysteresis between clockwise and counter-clockwise sweeps is most pronounced at 25 K and diminishes at elevated temperatures, reflecting temperature-dependent magnetic coercivity.

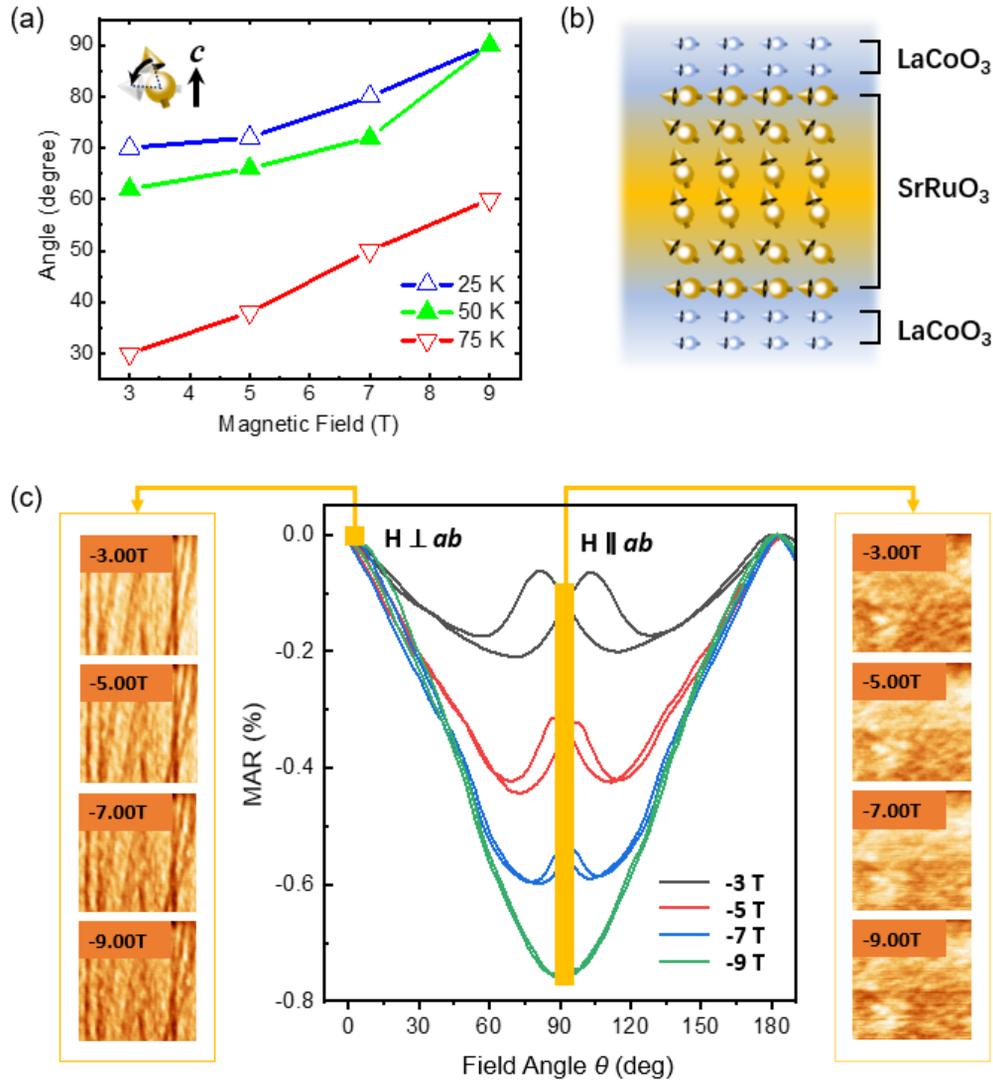

**Figure 4** Field-driven evolution of spin configuration and its correlation with magnetoresistance anisotropy. (a) Magnetic field dependence of the angle corresponding to minimum resistivity ($\theta_{min}$) at 25 K (blue triangles), 50 K (green triangles), and 75 K (red triangles). The increasing trend of $\theta_{min}$ with field strength indicates field-driven spin reorientation toward the film plane. The inset illustrates the measurement geometry with the magnetic field rotating in the a-c plane. (b) Schematic illustration of the proposed noncollinear spin configuration in the SRO layer. The spin orientation evolves from perpendicular to the film plane (bulk-like, orange region) to in-plane (interface region, yellow) due to interfacial exchange coupling with LCO layers, which exhibit in-plane magnetic anisotropy (blue arrows). This depth-dependent spin reorientation creates a canted spin texture within the SRO layer. (c) Magnetic angle-resolved (MAR) curves at 25 K for various magnetic fields (−3 T to −9 T) as a function of field angle $\theta$. Left panel: MFM images acquired at $\theta = 0°$ (H ⊥ ab) showing stripe domains that gradually diminish with increasing field strength. Right panel: MFM images acquired at $\theta = 90°$ (H // ab) showing absence of stripe domains at all measured fields. The correlation between stripe-domain formation (vertical field) and enhanced resistivity, as well as uniform magnetization (in-plane field) and suppressed resistivity, demonstrates the dominant role of magnetic domain structure in governing the magneto-transport properties.

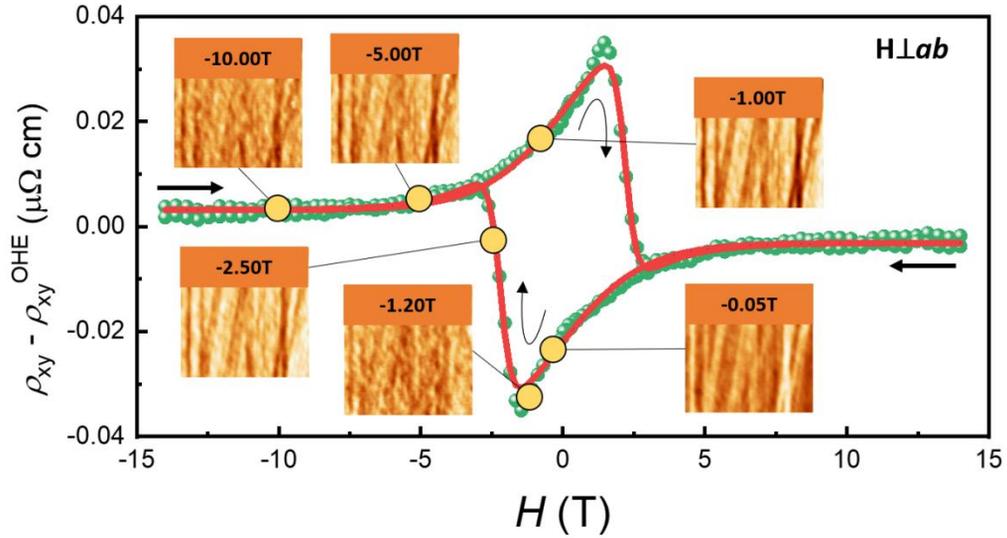

**Figure 5** Hall resistivity and magnetic domain evolution under vertical magnetic fields. Field-dependent Hall resistivity (after subtracting the ordinary Hall contribution, $\rho_{xy} - \rho_{xy}^{OHE}$) measured at 25 K with the magnetic field applied perpendicular to the film plane (H $\perp$ ab). Green symbols represent experimental data, while the thick red curve shows the fit using a two-channel anomalous Hall effect (AHE) model. The arrows indicate the magnetic field sweep direction. MFM images acquired at representative magnetic fields (−10.00 T, −5.00 T, −2.50 T, −1.20 T, −1.00 T, and −0.05 T) are overlaid, with yellow circles marking the corresponding positions on the Hall curve. Notably, the stripe-like magnetic domains are most pronounced at −0.05 T and −1.00 T, but become significantly suppressed near −1.2 T—coinciding with the peak feature in the Hall resistivity. This anticorrelation between stripe-domain visibility and Hall resistivity magnitude suggests that the peak structure originates from spin reorientation-induced transport anisotropy rather than domain-wall scattering. The deviation between the experimental data and the two-channel model fit around the peak and valley regions indicates contributions from the noncollinear spin texture that are not captured by conventional AHE models. The absence of circular skyrmion-like features in all MFM images confirms the suppression of topological Hall effects in these superlattices.